\documentclass[RNAAS]{aastex62}

\begin{document}

\title{Asteroseismic ages of red-giant stars from grid-based modelling: the impact of systematics in effective temperature or metallicity}

\correspondingauthor{Saskia Hekker}
\email{hekker@mps.mpg.de}

\author[0000-0002-1463-726X]{Saskia Hekker}
\affiliation{Max Planck Institute for Solar System Research, G\"ottingen, Germany}
\affiliation{Stellar Astrophysics Centre, Aarhus University, Aarhus, Denmark}

\author[0000-0002-6163-3472]{Sarbani Basu}
\affiliation{Yale University, New Haven, USA}


\section{} 
An increasingly popular method to determine stellar ages of red-giant stars for the purpose of Galactic archaeology is asteroseismic\footnote{Asteroseismology is the study of the internal structures of stars through their global oscillations.} grid-based modelling (GBM). In GBM, observed parameters are compared to those obtained from a grid of stellar evolution models to obtain stellar parameters such as mass, radius and age. In asteroseismic GBM of red-giant stars with solar-like oscillations the large frequency separation ($\Delta\nu$) and the frequency of maximum oscillation power ($\nu_{\rm max}$) are commonly used asteroseismic observables, in addition to the usual spectroscopic parameters effective temperature ($T_{\rm eff}$) and metallicity ([Fe/H]). Different types of observations are required to obtain the observed parameters: $\Delta\nu$ and $\nu_{\rm max}$ require timeseries data, while a single epoch observation is sufficient to determine $T_{\rm eff}$ and [Fe/H]. 

The precision with which $\Delta\nu$ and $\nu_{\rm max}$ can be determined largely depends on the length of the timeseries data (assuming the stars are bright enough that the oscillations can be detected). An increase in the timespan of the data and hence, the precision of the asteroseismic parameters is often obtained at the cost of number of stars observed. Different space mission have made different choices. The \textit{Kepler} mission provided the longest ($\sim$ 4 years long) timeseries data of red giants with solar-like oscillations currently available. At the other extreme, the TESS mission is providing $\sim$27-day to about 1-year long timeseries for tens of thousands of oscillating red giants. The question that we aim to answer here is: with what precision should [Fe/H] and $T_{\rm eff}$ be obtained to derive stellar ages of red-giant stars through asteroseismic GBM given the precision of $\Delta\nu$ and $\nu_{\rm max}$ that we can expect from the TESS data?

To answer this question, we investigate the impact of uncertainties and biases (i.e., systematic errors) in $T_{\rm eff}$ and [Fe/H] on the determined ages of red-giant stars using asteroseismic GBM, given the precisions to which $\Delta\nu$ and $\nu_{\rm max}$ can be obtained with timeseries data of $\sim 50$ and $\sim 400$ day length. For this investigation we use the APOKASC sample of $\sim6600$ red-giant stars \citep{pinsonneault2018,elsworth2019} and perform asteroseismic GBM on these red-giant stars using the MPS-SAGE code \citep[for details see][and Serenelli et al. in prep.]{hekker2014}.  We estimated the uncertainties to be $\sigma_{\Delta\nu}=0.1$~$\mu$Hz and  $\sigma_{\nu_{\rm max}}=1$~$\mu$Hz for a 50-day dataset and $\sigma_{\Delta\nu}=0.05$~$\mu$Hz and $\sigma_{\nu_{\rm max}}=0.5$~$\mu$Hz for a 400-day dataset as per \citet{hekker2012}.  From now on we refer to these datasets as the 50-day and 400-day dataset.

\begin{figure}
\begin{center}
\includegraphics[width=0.9\textwidth]{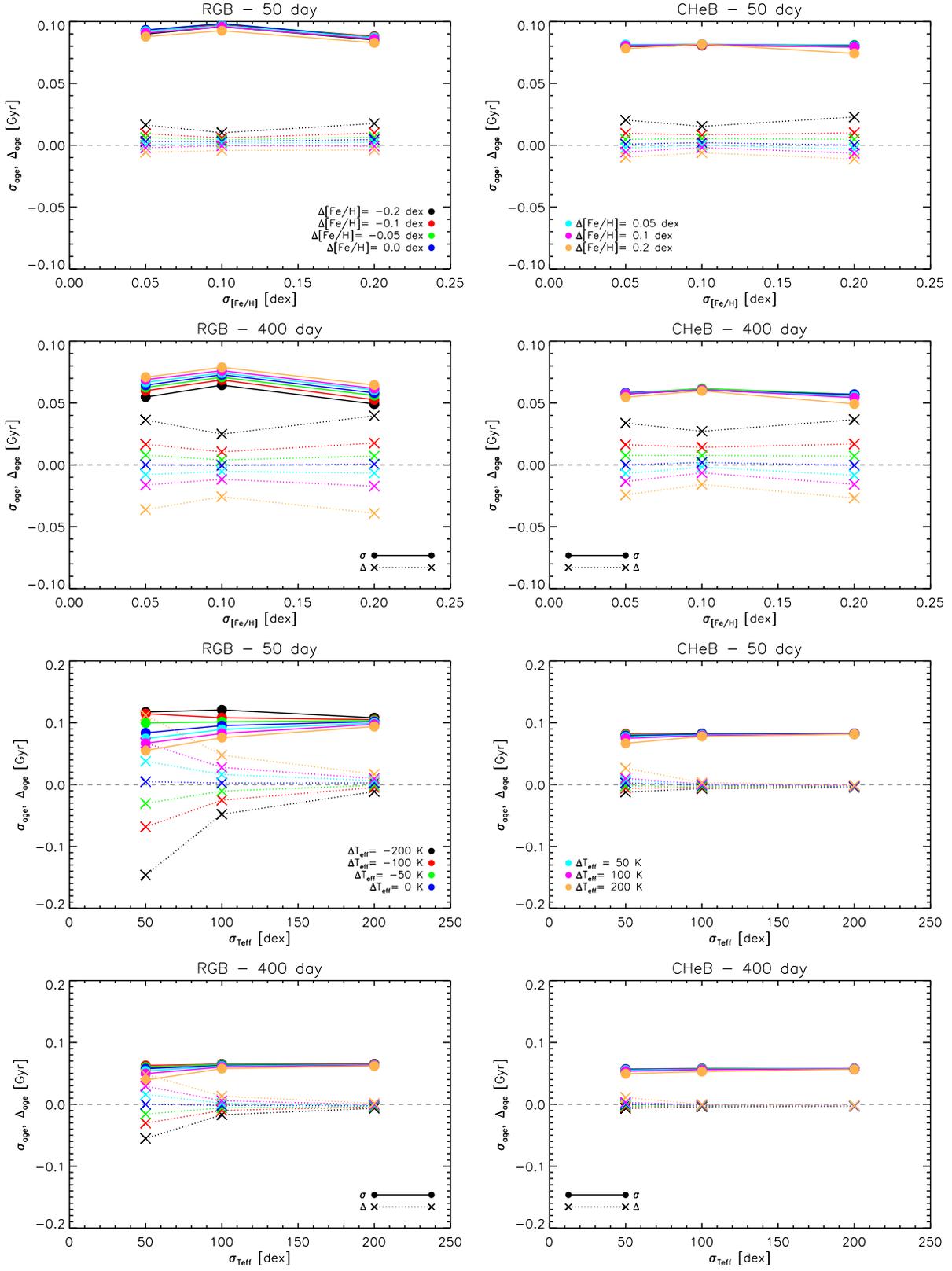}
\caption{Uncertainties $\sigma_{\rm age}$ (dots with solid lines) and biases $\Delta_{\rm age}$ (crosses with dotted lines) in age estimates obtained from asteroseismic GBM as a function of uncertainty in [Fe/H] (top 4 panels) or $T_{\rm eff}$ (bottom 4 panels). The biases in [Fe/H] and $T_{\rm eff}$ are colour-coded as per the legends. Panel headers indicate the length of the dataset and the evolutionary state of the stars presented. The gray dashed lines indicate zero.}
\end{center}
\label{figure}
\end{figure}

For both the 50-day and 400-day datasets we compute ages and uncertainties in ages for $T_{\rm eff}$ and [Fe/H] values assuming the following uncertainties:
\begin{eqnarray}
\sigma_{T_{\rm eff}}=[50, 100, 200]\textrm{ K,}\\
\sigma_{\textrm{[Fe/H]}} = [0.05, 0.1, 0.2]\textrm{ dex.}
\end{eqnarray}
We additionally assume the following biases or systematic errors:
\begin{eqnarray}
\Delta_{T_{\rm eff}} = [-200,-100,-50,0,50,100,200]\textrm{ K,}\\
\Delta_{\textrm{[Fe/H]}} = [-0.2,-0.1,-0.05,0.0,0.05,0.1,0.2]\textrm{ dex.}
\end{eqnarray}
We note here that the tests for $T_{\rm eff}$ and [Fe/H] are performed independently, with the value of the second parameter fixed to the value as provided by APOKASC \citep{pinsonneault2018}.

In Fig.~\ref{figure}, we present the resulting uncertainties and biases in ages for red-giant stars given the uncertainties and biases in [Fe/H] (top 4 panels) and $T_{\rm eff}$ (bottom 4 panels). The biases are computed with respect to the 400-day dataset with no bias (dark blue) and smallest uncertainty. In most cases presented here the value of $\sigma_{\rm age}$ shows only weak, or no, dependence on the uncertainties and biases in [Fe/H] and $T_{\rm eff}$ ingested in this study. The largest deviations are present in the 50-day dataset with $T_{\rm eff}$ biases. Furthermore, in most cases  $\Delta_{\rm age} < \sigma_{\rm age}$ and hence a potential bias is incorporated in the uncertainties. Only in case a large ($>100$~K) $T_{\rm eff}$ biases is accompanied by a small (50 K) uncertainties in $T_{\rm eff}$ on the red-giant branch, $\Delta_{\rm age}$ is of the order of or larger than $\sigma_{\rm age}$ and in these cases we expect an impact of this bias on the resulting age estimate. These results are in line with the  robustness of main-sequence and subgiant solar-like oscillators to biases and uncertainties in $T_{\rm eff}$ and [Fe/H] presented by \citet{bellinger2019} .

\acknowledgments

\end{document}